\documentclass[%
 reprint,
 amsmath,amssymb,
prb,
]{revtex4-1}

\usepackage{graphicx}
\usepackage{dcolumn}
\usepackage{bm}

\begin{document}

\preprint{APS/123-QED}

\title{Large spin current generation by the spin Hall effect in mixed crystalline phase Ta thin films}

\author{Akash Kumar}
\author{Rajni Bansal}
\author{Sujeet Chaudhary}%
\author{Pranaba Kishor Muduli}
 
 \email{muduli@physics.iitd.ac.in}
\affiliation{%
 Thin Film Laboratory, Department of physics,\\
 Indian Institute of Technology Delhi, New Delhi, India--110016 
}%

\date{\today}

\begin{abstract}
Manipulation of the magnetization in heavy-metal/ferromagnetic bilayers via the spin-orbit torque requires high spin Hall conductivity of the heavy metal.
We measure inverse spin Hall voltage using a co-planar wave-guide based broadband ferromagnetic resonance set-up in Py/Ta system with varying crystalline phase of Ta. 
We demonstrate a strong correlation between the measured spin mixing conductance and spin Hall conductivity with the crystalline phase of Ta thin films. We found a large spin Hall conductivity of $-2439~(\hbar/e)~\Omega^{-1}$cm$^{-1}$ for low-resistivity (68~$\mu\Omega$--cm) Ta film having mixed crystalline phase, which we attribute to an extrinsic mechanism of the spin Hall effect. 
\begin{description}
\item[Keywords]
Spin Hall effect, Ta thin films, Spin mixing conductance, Inverse spin Hall effect
\end{description}
\end{abstract}

\pacs{Valid PACS appear here}
\maketitle



Spin Hall effect (SHE) can be used to produce a pure transverse spin current density ($J_{\rm s}$) from a longitudinal electrical current density ($J_{\rm e}$) in heavy metals.~\cite{Dyakonov1971,Hirsch1999}  The pure spin current can be measured using the reciprocal effect, \textit{i.e.}, the inverse spin Hall effect (ISHE) employing a transverse charge current created from the pure spin current. The spin current can generate a current-induced spin-orbit torque (SOT) in heavy metal/ferromagnet (HM/FM) heterostructure for potential application in the efficient manipulation of magnetization at the nanoscale.~\cite{Liu2011,Liu2012}
With sufficiently strong SOT, it is possible to excite magnetization to auto-oscillation for radio frequency generation application~\cite{Liu2012PRL,VEDemidov2012,Dhananjay2017} or switch the magnetization, move domain walls or skyrmions for efficient memory applications.~\cite{Liu2012,Miron2014,Liu2012PRL,Debanjan2014} 

For realizing these applications, a large spin Hall angle, $\theta_{\rm SH}$ defined as the ratio of the spin current density to the charge current density is desirable. While the value of $\theta_{\rm SH}$  in most commonly investigated metal Pt is $\theta_{\rm SH}\leq~0.12$,~\cite{Ando2008,Liu2011,Azevedo2011,sinova2015rmp}
recent results show relatively higher spin Hall angle of $|\theta_{\rm SH}| \leq 0.25$ in Ta,\cite{Morota2011, Liu2012,Hao2015,Allen2015,Kim2015,Nimi2015,Qiu2014,PDeorani2013,Hahn2013,Velez2016,Emori2013} and of the order of $|\theta_{\rm SH}|\leq0.50$ for W.~\cite{CFPai2012,Demasius2016Ncomm,Hao2015}
However, these higher values of $\theta_{\rm SH}$ in Ta and W are so far reported in very high resistive phase of these materials, which limits several applications that require a charge current to flow in the HM.

In this work, we report a strong correlation of spin Hall angle with the crystalline phase of Ta thin films in Py/Ta bilayers. The crystalline phase of Ta films is varied by controlling growth rate in sputtering. We develop and demonstrate a simple method for measurement of ISHE using a broadband ferromagnetic resonance (FMR) set-up without involving micro-fabrication. We show that the voltage measured in our optimized set-up primarily arises from ISHE by using out-of-plane angle dependence and radio frequency (RF) power dependence, which rules out voltage signal due to other galvanomagnetic effects such as anisotropic magneto-resistance (AMR) and anomalous Hall effect (AHE). We find a higher spin mixing conductance and spin Hall conductivity ($-2439~(\hbar/e)~\Omega^{-1}$cm$^{-1}$) for \textit{low resistivity} Ta having mixed crystalline phase, which is promising for applications. The large spin Hall conductivity for mixed crystalline phase Ta is consistent with the extrinsic mechanism of spin Hall effect.


The Py($t_{\rm Py}$ nm)/Ta(20 nm) bilayer thin films are prepared on Si substrates using DC-magnetron sputtering at a working and base pressure of $2\times10^{-3}$ and $3\times10^{-6}$ Torr, respectively. We first studied single layer Ta thin films with different growth rates by varying the DC-sputtering power. Subsequently, Py($t_{\rm Py}$ nm)/Ta(20 nm) bilayer thin films were prepared with varying thickness of Py, $t_{\rm Py}$ and growth rate of Ta. The Ta thickness was kept fixed at 20~nm. Before the deposition of the different layers, pre-sputtering of the targets was performed for 10 min with a shutter. Crystallographic properties of films were determined using X-ray diffraction (XRD) while the thicknesses and interface/surface roughness were determined from X-ray reflectivity (XRR) technique using a PANalytical X'Pert diffractometer with Cu-K$_{\alpha}$ radiation. 
The  XRR data (not shown) was fitted using the recursive theory of Parratt.~\cite{parratt1954}
From XRR fitting the surface and interface roughness were found to be $<$0.5~nm.  

\begin{figure}[t]
\centering
\includegraphics[width=9cm]{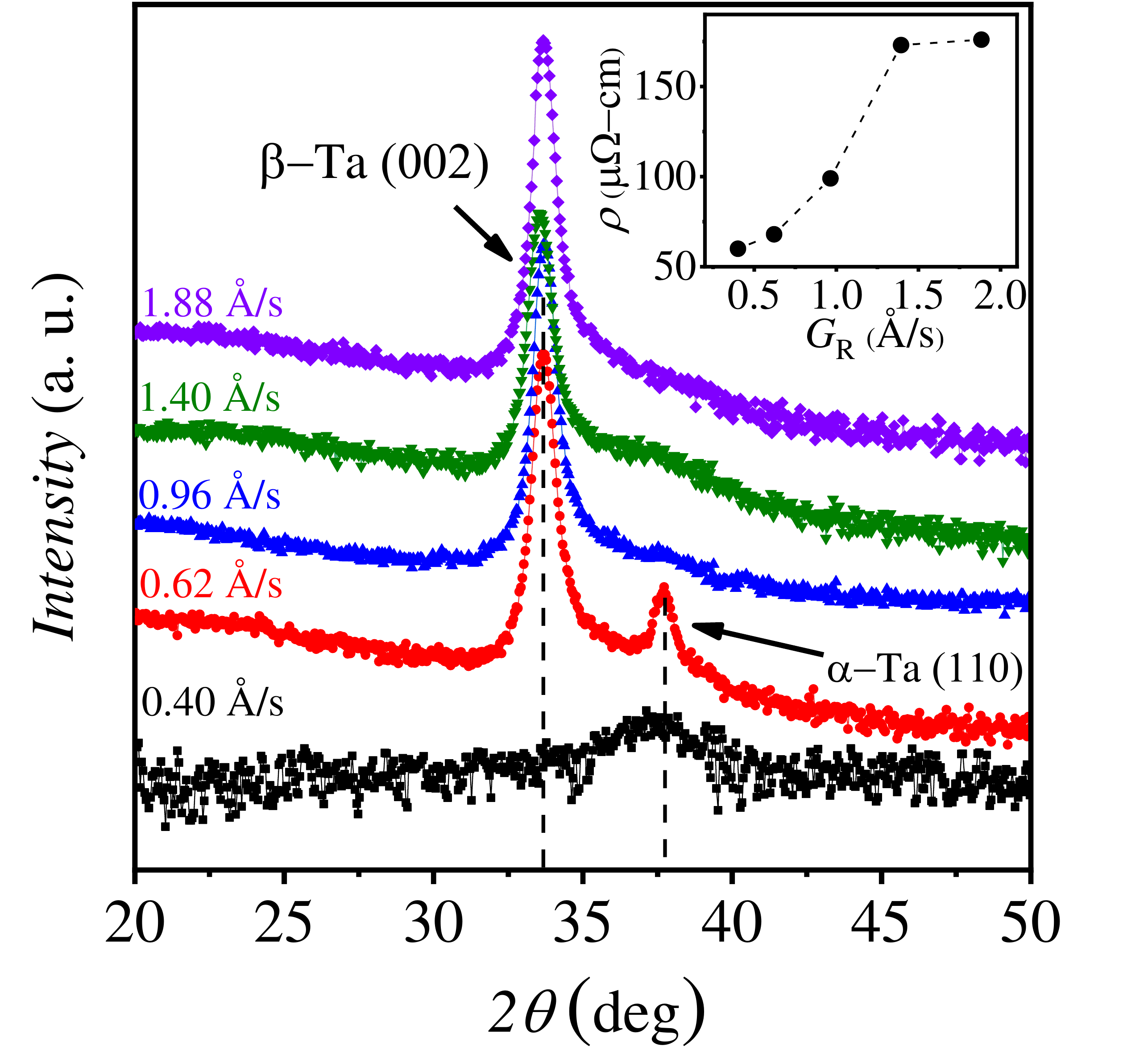}\\
\caption{(a) X-ray diffraction data for Ta thin films grown at different growth rates ($G_{R}$). The plots are shifted for clarity. The inset shows the resistivity versus $G_{R}$ for Ta thin films having thickness of 20~nm.}\label{XRD_Ta}
\end{figure}

Ferromagnetic resonance (FMR) measurements are carried out for excitation frequencies of 4--12 GHz at room temperature. We use a co-planar waveguide (CPW) based broad-band FMR set-up.~\cite{bansal2018apl}
For a fixed excitation frequency of microwave field, external magnetic field ($H$) is swept for the resonance condition. 
The ISHE measurements are performed on $4\times3~$mm$^{2}$ samples by measuring voltage signal at the edge of the samples by fabricating 100~$\mu$m-thick Cu contact pads. This geometry allows us to measure ISHE signal in our samples when the film side is facing the CPW.

\begin{figure} [t!]
\centering
  \includegraphics[width=8.5cm]{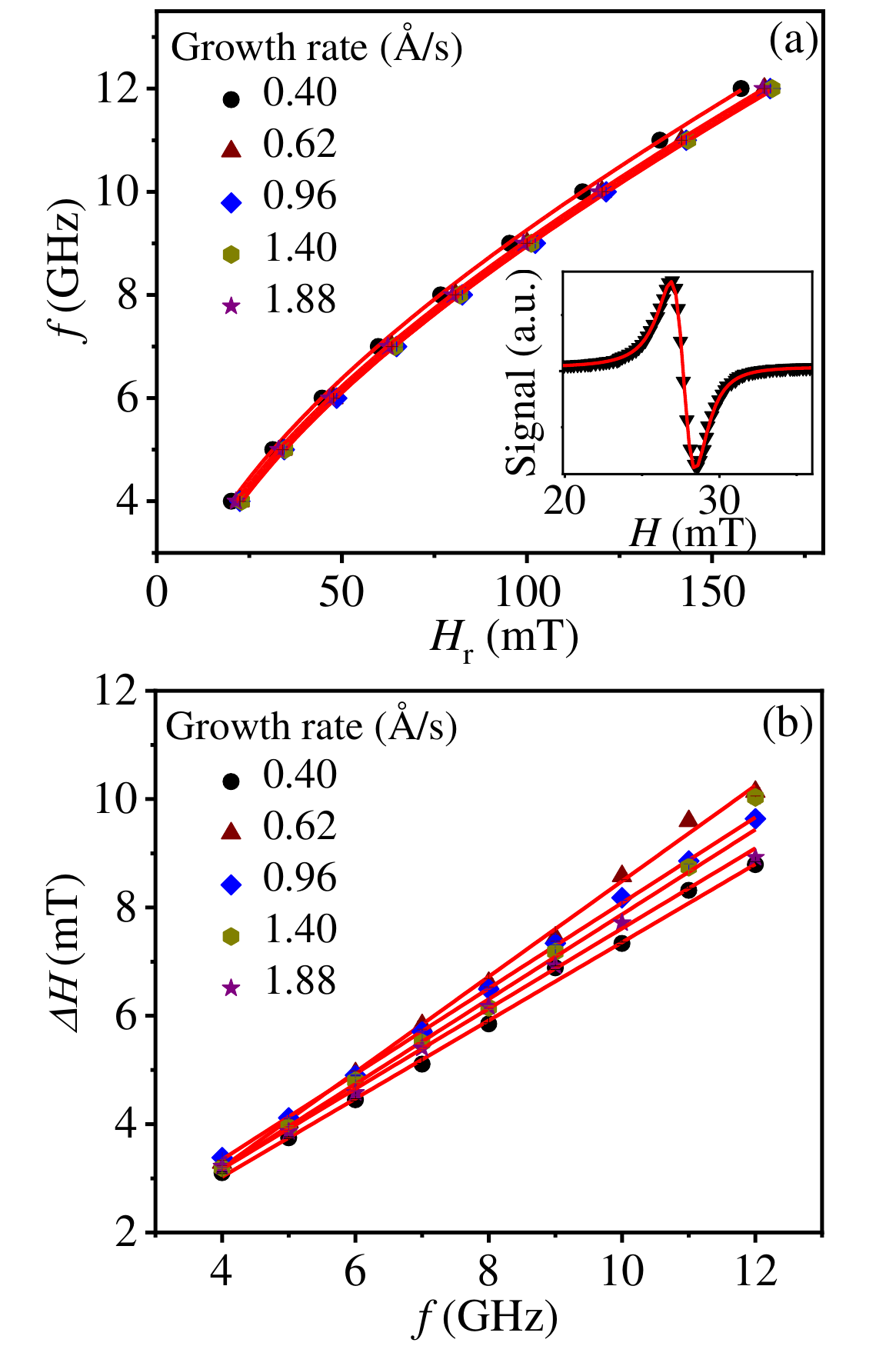}\\
  \caption{(a) Frequency ($f$) \textit{vs.} resonance field ($H_{r}$) and (b) Linewidth ($\Delta H$) vs. frequency, $f$ for Py (30 nm)/Ta (20 nm) with varying phase of Ta obtained by varying the growth rates. The inset in (a) shows an example FMR spectra at 5 GHz for $G_{R}=0.62~\rm \AA$/s. The symbols are measured data while solid lines are fits.}\label{FMR}
\end{figure}
Figure~\ref{XRD_Ta} shows the XRD spectra for single layer 50~nm-thick Ta thin films prepared at different growth rates by varying sputtering power in DC magnetron sputtering. 
A broad diffusive peak of $\alpha$-phase of Ta centered around 2$\theta = 38.0^\circ$ is observed for thin films grown at the lowest growth rate of 0.40~\AA/s. This peak corresponds to (110) reflection of $\alpha-$Ta. Bragg peaks corresponding to (002) $\beta$-Ta and (110) $\alpha$-Ta are observed for growth rates between 0.62~\AA/s and 1.4~\AA/s, respectively, which suggest the growth of a mixed ($\alpha$+$\beta$) phase of Ta. However, an oriented $\beta$-phase of Ta is observed for growth rate higher than 1.4 \AA/s with Bragg reflections at 2$\theta$ = 33.6$^\circ$ corresponding to $\beta$-Ta (002) reflection. Inset of Fig.~\ref{XRD_Ta} shows the resistivity measurements on 20 nm Ta thin films grown with varying growth rate, measured using Van der Pauw method. The samples with pure $\beta$-phase of Ta shows a higher resistivity of about 180~$\mu\Omega$--cm which is in agreement with literature.~\cite{Liu2012}
For the $\alpha$-phase the resistivity is found to be around 60~$\mu\Omega$--cm.

The measured FMR data are shown in Fig.~\ref{FMR} for Py(30~nm)/Ta(20~nm) bilayer structure, where Ta is grown at a growth rate of 0.62 $\rm \AA$/s at which a mixed ($\alpha$+$\beta$) phase of Ta is formed. The raw FMR spectra are fitted with the sum of derivative of symmetric and asymmetric Lorentzian functions:~\cite{Woltersdorf2004}  

\begin{equation}\label{Vdc}
\begin{split}
V_{\rm dc}= -2V_{\rm sym} \frac{\Delta H^{2} \small(H-H_{\rm r})}{(\Delta H^2+\small(H-H_{\rm r})^2)^2}\\+V_{\rm asym} \frac{ \Delta H\small(\Delta H^2 -\small(H-H_{\rm r})^2)}{(\Delta H^2+\small(H-H_{\rm r})^2)^2},
\end{split}
\end{equation}

where $H$, $\Delta H$, and $H_{\rm r}$ are the measured field, FMR linewidth (half width at half maximum; HWHM) and resonance field, respectively.  $V_{\rm sym}$ and $V_{\rm asym}$ are the symmetric and asymmetric amplitudes of the voltage signal, respectively.  An example of FMR spectra with the fitting is shown in the inset of Fig.~\ref{FMR}(a).

From the fittings of FMR spectra, the linewidth $(\Delta H)$ and the resonance field ($H_{\rm r}$) are determined. The $H_{\rm r}$ as a function of frequency ($\it f$) is shown in Fig.~\ref{FMR}(a), which was fitted with Kittel equation:~\cite{Kittle1948}
\begin{equation}\label{Kittel}
f=\frac{\gamma}{2\pi}[(H_{\rm r}+H_{\rm k})(H_{\rm r}+H_{\rm k}+4\pi M_{\rm eff})]^{1/2},
\end{equation}
 
where, $M_{\rm eff}$ is the effective saturation magnetization and $H_{\rm k}$ is the anisotropy field. Here, $\gamma$=1.85$\times10^{2}$~GHz/T is the gyromagnetic ratio. 

\begin{figure}[t]
\centering
  \includegraphics[width=8.5cm]{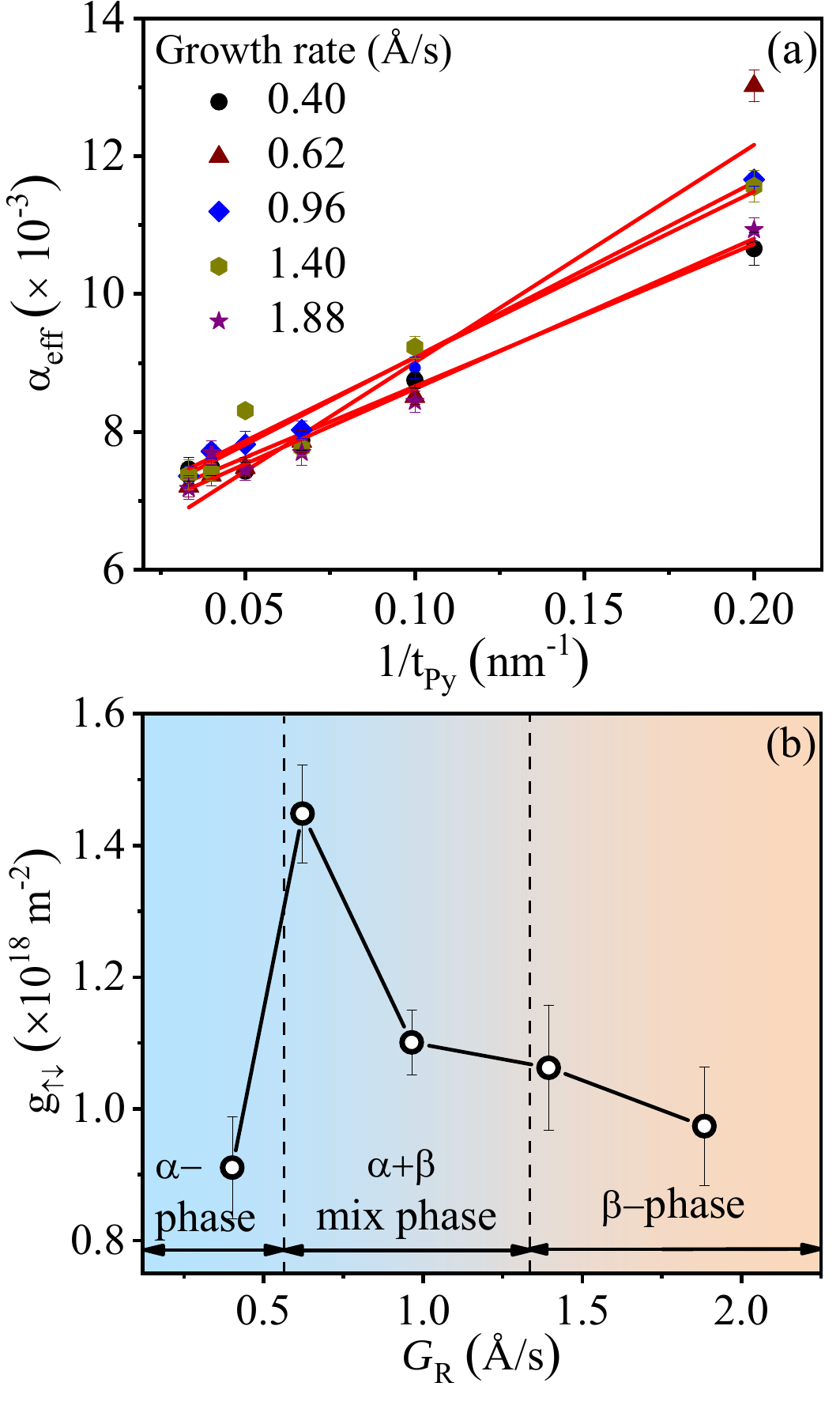}\\
  \caption{(a) Effective damping constant ($\alpha_{\rm eff}$) vs. inverse of Py thickness, for different crystalline phases of Ta. (b) Variation of $\mathrm{g}_{\uparrow \downarrow}$ with growth rate $G_{\rm R}$ of Ta.}\label{SMC}
\end{figure}

The Gilbert damping parameter, $\alpha $ was calculated from the slope of the $\Delta H$ vs. $f$ [Fig.~\ref{FMR}(b)] by fitting with following equation:
\begin{equation}\label{eq;deltah}
\Delta H=\frac{2\pi\alpha_{\rm eff} f}{\gamma}+\Delta H_{\rm 0},
\end{equation}
 where $\Delta H_{\rm 0}$ is inhomogeneous line broadening, which is related with the film quality. In our experimental results [Fig.~\ref{FMR}(b)], the $\Delta H$ vs. $f$ shows a linear behavior indicating the intrinsic origin of damping parameter observed in our Py/Ta bilayers thin films. We have also observed very small value of $\Delta H_{\rm 0}$ ($< 1$~mT), which further confirms the high-quality of these thin films. For quantifying spin pumping for different Ta crystalline phase, we have performed Py thickness dependence of $\alpha_{\rm eff}$ and $M_{\rm eff}$ for varying crystalline phase of Ta. Figure~\ref{SMC}(a) shows damping parameter vs. inverse of Py thickness for the different crystalline phase of Ta thin films.  We then calculate the spin mixing conductance, $\mathrm{g}_{\uparrow \downarrow}$ which is an important parameter that determines the spin pumping efficiency. According to the theory of spin pumping,~\cite{Tserkvovyak2002}  
\begin{equation}\label{eq:spump}
\alpha_{\rm eff} = \alpha_{0}+g\mu_{0}\mu_{B}\frac{\textsl{g}_{\uparrow\downarrow}}{4\pi M_{\rm s}}\frac{1}{t_{\rm FM}},
\end{equation}

where, $g$ and $\mu_{B}$ are Land\'e $g$-factor and Bohr magneton, respectively. We have calculated $\mathrm{g}_{\uparrow \downarrow}$ by fitting Gilbert damping parameter $(\alpha_{\rm eff})$ versus inverse of Py thickness with above equation as shown in Fig.~\ref{SMC}(a). We used $g=2.1$ for Ni$_{80}$Fe$_{20}$ for calculating $\textsl{g}_{\uparrow\downarrow}.$~\cite{Shaw2013} In Eq~(\ref{eq:spump}), we neglected the spin back flow, since the Ta thickness we used is quite large compared to reported spin diffusion length of Ta.~\cite{Liu2012,Emori2013,Wang2014,Kim2015,Allen2015,Nilamani2016} 

\begin{figure}[t]
  \includegraphics[width=9cm]{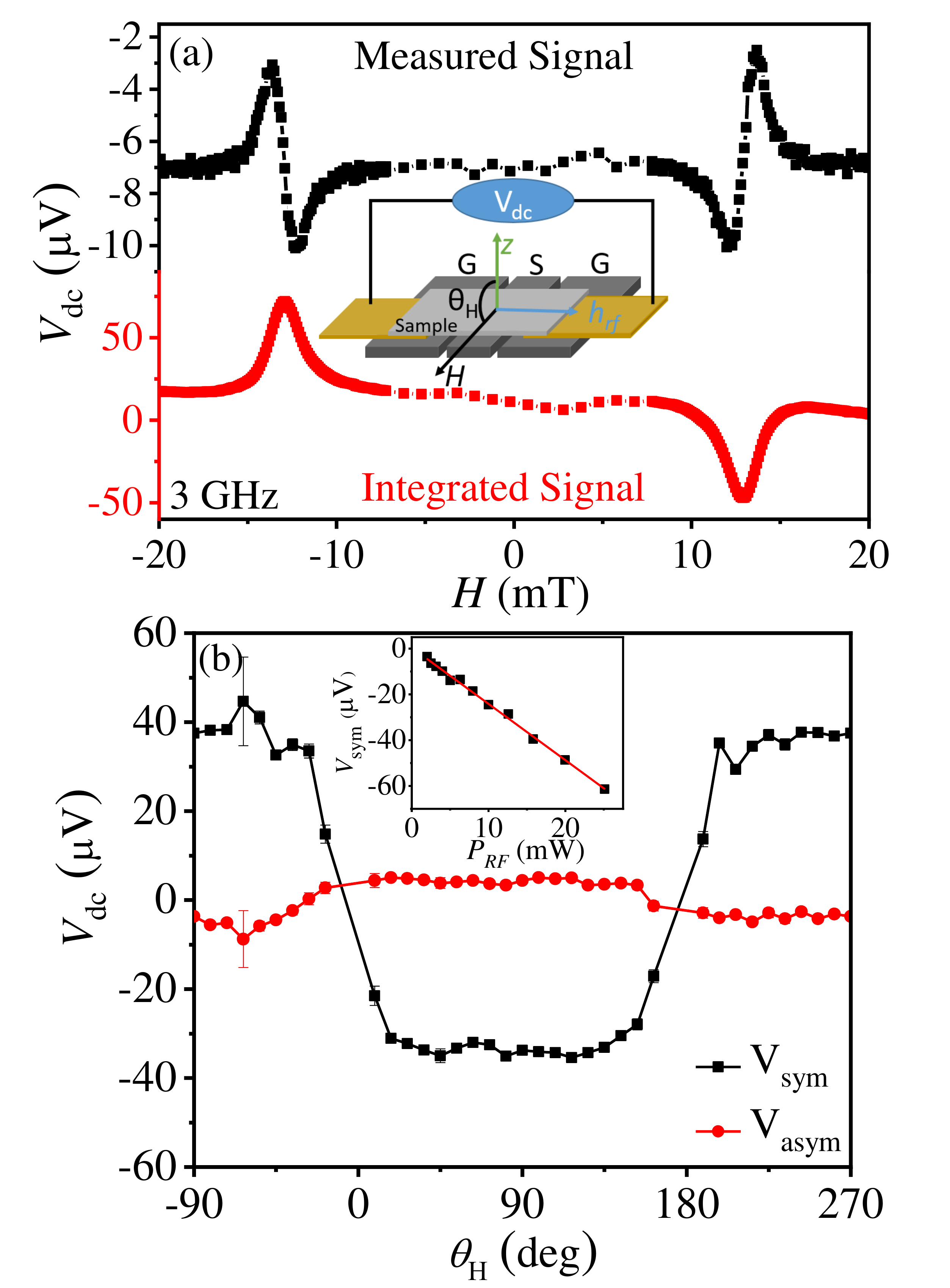}\\
  \caption{(a) Measured and integrated ISHE signal at 3 GHz, for Py(30 nm)/Ta(20 nm) with $G_{\rm R}=0.62~\AA$/s. The inset shows the schematic of the ISHE voltage measurement geometry. (b) Out-of plane ($\theta_{H}$) ISHE measurements with magnetic field applied out-of the film plane for 2~GHz of excitation frequency at  RF power of 15.85~mW. Inset shows $V_{\rm sym}$ versus $P_{\rm RF}$ for 3~GHz. The solid line is a linear fit.}\label{ISHE}
\end{figure}

Figure~\ref{SMC}(b) shows the value of $\textsl{g} _{\uparrow\downarrow}$ with varying growth rate of Ta thin films. Interestingly, the highest value of $\textsl{g} _{\uparrow\downarrow}$ is observed for the mixed phase of Ta. In a recent study, we showed that the spin current efficiency is maximum for the mixed phase Ta using an optical technique.~\cite{Rajni2017} Thus, the higher value of $\textsl{g} _{\uparrow\downarrow}$ for mixed phase Ta is consistent with this earlier study. The spin mixing conductance,  $\textsl{g} _{\uparrow\downarrow}$ determines the amount of spin current injected to the non-magnetic Ta layer. A variation of $\textsl{g} _{\uparrow\downarrow}$ with crystalline phase, imply a change of effective spin current injected to the Ta layer. Hence, a correlation between $\textsl{g} _{\uparrow\downarrow}$ and the inverse spin Hall voltage is expected.
For this, we measured ISHE in these bilayers as a function of the crystalline phase of Ta thin films. The upper panel in Fig.~\ref{ISHE}(a) shows an example of ISHE voltage signal observed for the Py/Ta thin film for the growth rate of 0.62 $\rm \AA$/s. 

In our measurement, we have used a field-modulation method to enhance the sensitivity.~\cite{tiwari2016apl} In this method, the static field is modulated with a small ac field (98~Hz) of magnitude 0.5~mT, produced by a pair of Helmholtz coils. These coils are powered by the lock-in amplifier, which also measures the voltage across the sample.
Thus, the field modulation method measures essentially the derivative signal. However, the most reported literature on ISHE uses amplitude modulation of RF signal.~\cite{Mosendz2010,Mosendz2010PRL,Ando2011,PDeorani2013,Surbhi2017}
Hence, in the lower panel of Fig.~\ref{ISHE}(a), we show the integrated ISHE signal for better comparison with the literature. The measured signal in our system may consist of ISHE in the Ta layer, and the AMR or AHE of the Py layer. The AMR or AHE is often assumed to produce an asymmetric Lorentzian shape  while the ISHE is assumed to produce a symmetric Lorentzian shape~\cite{Mosendz2010PRL,Ando2011,Ando2012NC,Bai2013PRL,Iguchi2017} 
so that the measured data is a sum of symmetric and asymmetric Lorentzian functions. Our measured ISHE spectra are symmetric in shape and changes sign with inversion of magnetic field direction which indicates that the voltage signal we measure may be primarily due to ISHE.~\cite{Mosendz2010,Mosendz2010PRL,Ando2011,PDeorani2013}

To further verify that the measured signal is indeed from the ISHE, we measured the voltage in our samples by changing the direction of magnetic field out-of-the film plane. The measurement geometry is shown in the inset of Fig.~\ref{ISHE}(a). Here, the out-of-plane angle ($\theta_{\rm H}$) is measured from the \textit{z}-axis, so that $\theta_{H}=0^\circ$ corresponds to the out-of-plane direction. Figure~\ref{ISHE}(b) shows the variation of symmetric and asymmetric voltage components with varying $\theta_{\rm H}$. According to Lustikova \textit{et. al.}, the asymmetric component ($V_{\rm asym}$) can arise due to the AMR and AHE while the symmetric component ($V_{\rm sym}$) arises due to ISHE, as well as AMR and AHE.~\cite{Lustikova2015} 
In our measurements, we found $V_{\rm asym}<<V_{\rm sym}$ for the entire range of $\theta_{H}$. Also, the observed angular dependence of $V_{\rm asym}$ is not analogous to the analysis presented by Lustikova \textit{et al.}.~\cite{Lustikova2015}
This indicates that the voltage signal measured in our set-up is primarily due to the ISHE. Furthermore, the $V_{\rm sym}$ increases linearly with radio frequency(RF) power as shown in the inset of Fig.~\ref{ISHE}(b), which is also consistent with ISHE.~\cite{Ando2011}


Based on the above observations, we took the symmetric signal as ISHE ($V_{\rm sym} = V_{\rm ISHE}$) for calculating the spin Hall angle ($\theta_{\rm SH}$) of Ta. The spin Hall angle relates to the ISHE voltage in the following manner:~\cite{Ando2010,Mosendz2010,PDeorani2013,Wang2014,Surbhi2017}
\begin{equation}\label{VISHE}
V_{\rm ISHE}=\zeta \theta_{\rm SH}LR\lambda_{\rm Ta}\tanh\left( \frac{d_{\rm Ta}}{2\lambda_{\rm Ta}}\right)\times J^0_{\rm s}
\end{equation}
 
where, $V_{\rm ISHE}$ is the ISHE signal induced by spin pumping, $R$ is the sample resistance measured from Py/Ta samples, $L$ is the length of the sample, $d_{\rm Ta}$ and $\lambda_{\rm Ta}$ are thickness and spin diffusion length of Ta thin film, respectively. The spin diffusion length of Ta thin films is taken to be 2.47~nm measured in similar bilayer structures.~\cite{Nilamani2016} As a first approximation, we neglect the possible variation of $\lambda_{\rm Ta}$  with the crystalline phase of Ta. This assumption is very much valid in our case as the thickness of Ta is very large compared to the reported values of the spin diffusion length of Ta, which is of the order of 0.4--3~nm~\cite{Liu2012,Emori2013,Wang2014,Kim2015,liu2015apl, Allen2015,Nilamani2016,beherarsca2017} so that $\tanh\left( \frac{d_{\rm Ta}}{2\lambda_{\rm Ta}}\right)\approx 1$ in the above equation.
The spin back flow is also negligible for the same reason \textit{i.e.,}  $d_{\rm Ta} \gg \lambda_{\rm Ta}$. 
$\zeta$ is the correction factor and comes from the fact that only a part of the sample contributes to the spin pumping and it depends upon the area of the sample above the signal line of CPW. The value of $\zeta$ can be calculated using the method discussed by P. Deorani \textit{et al.}~\cite{PDeorani2013} by noting that the width ($w$) of signal line is 185~$\mu m$ wide and distance between contact pads is 2~mm and spin wave propagation length of Ta is about 25 $\mu m$.~\cite{Kwon2013}

$J^0_{\rm S}$ is spin current density and can be defined as,
\begin{equation}\label{Js}
J^0_{\rm s}=\frac{2e}{\hbar}\times\frac{\hbar\omega}{4\pi} Re(\mathsf{g_{\uparrow\downarrow}}) \sin^2\theta_{\rm C}\times P, 
\end{equation}

where, $\omega=2\pi f$ is the angular frequency of microwave excitation, $Re(\mathsf{g_{\uparrow\downarrow}})$ is real part spin mixing conductance. $\theta_{\rm C}$ is the magnetization precession cone angle given by $\theta_{\rm C}=\frac{h_{\rm rf}}{2\Delta H}$, where $h_{\rm rf}$ is the strength of RF field experienced by the sample. This field is generated around the signal line as a result of RF current flow. The value of $h_{\rm rf}$ is obtained from Ampere’s law, $h_{\rm rf} = \frac{\mu_{\rm 0}}{2\pi w} \sqrt[]{\frac{P_{\rm RF}}{Z}}\log(1+\frac{w}{D})$, where $w$ and $D$ is the width of signal line and separation between signal line to sample (0.5 mm), $P_{\rm RF}$ is the applied RF power and $Z$ is the impedance of CPW \textit{i.e.,} 50~$\Omega$. $P$ is the ellipticity correction factor and arises from the ellipticity of the magnetization precession. As the magnetization precession in a magnetic thin film is not exactly circular, but follows an elliptical path due to strong demagnetizing fields. According to Ando \textit{et al.}\cite{Ando2009}, $P$  is given by,

\begin{equation}\label{P}
P= \frac{2\omega[M\gamma+\sqrt(M\gamma)^2+(2\omega)^2]}{(M\gamma)^2+(2\omega)^2}
\end{equation}

where, $M=4\pi M_{\rm s}$. As explained by Mosendz \textit{et al.}\cite{Mosendz2010}, the value of ellipticity correction factor $P$ changes by a multiple of more than 3 for frequency range 2 to 8 GHz and becomes larger than 1 for frequencies higher than 10 GHz. This shows that magnetization precession trajectory is very elliptical for lower frequencies and DC voltage due to ISHE requires this correction factor. 
\begin{figure}[t]
\centering  \includegraphics[width=8.5cm]{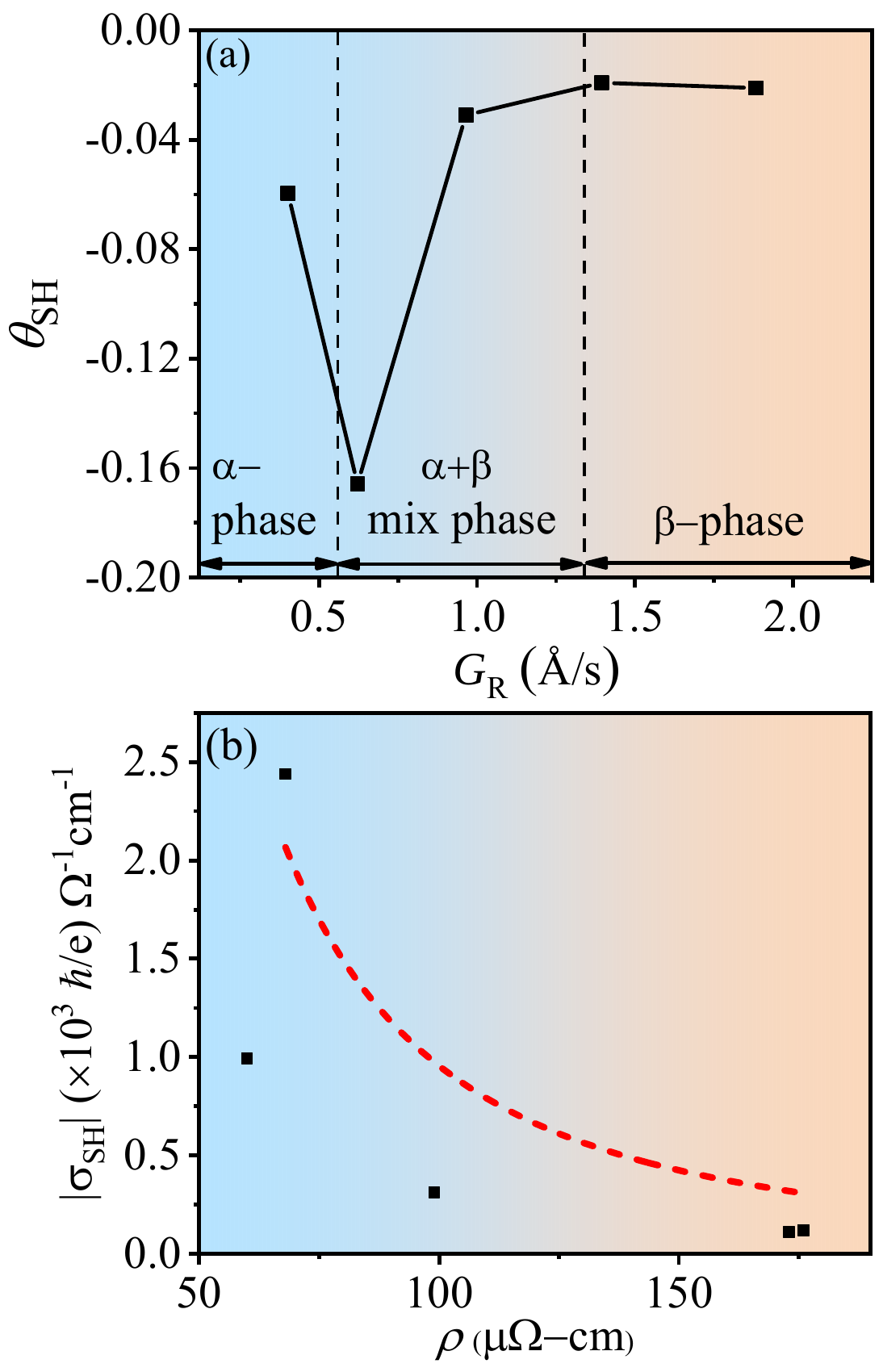}\\
\caption{(a) Measured Spin Hall angle ($\theta_{\rm SH}$) versus growth rate, $G_{R}$ of Ta for the Py (30 nm)/Ta (20 nm) bilayer thin films at 3 GHz FMR frequency. (b) Spin Hall conductivity ($\sigma_{\rm SH}$) as a function of resistivity ($\rho$) of Ta thin films.}\label{SHA}
\end{figure}
After considering all the required terms, the value of the spin Hall angle of the Ta thin films can be calculated using Eq.~(\ref{VISHE}).

We found the sign of spin Hall angle to be negative, which was confirmed by measuring a Py/Pt sample (not shown) for which the sign was found positive which is consistent with the literature. 
Figure~\ref{SHA} shows the variation of spin Hall angle with varying growth rate of Ta, measured at 3 GHz.  The results show a strong correlation between the spin Hall angle and crystalline phase of Ta. Furthermore, the behavior of spin Hall angle is similar to the variation of spin mixing conductance shown in Fig.~\ref{SMC}(b) indicating a strong correlation between the spin-mixing conductance and the inverse spin Hall effect. This correlation further confirms that spin rectification effects are negligible in our measurement, unlike a recent study where spin-mixing conductance and the inverse spin Hall effect were found to be uncorrelated due to the presence of spin rectification effects.~\cite{conca2017prb} 

Surprisingly,  it is observed that the low-resistivity Ta thin films with mixed phase show the highest value of spin Hall angle, $-0.16\pm 0.01$. The highest value of spin Hall angle reported in the literature for Ta is $-0.25$, which was for the high resistive $\beta$-phase of Ta.~\cite{Emori2013} In our case, we observed a lower spin Hall angle of $\theta_{\rm SH}\approx-0.02$ for the high resistive $\beta$-phase of Ta. We calculate the spin Hall conductivity, $\sigma_{\rm SH}$ using the following formula: 
\begin{equation}\label{Js}
\sigma_{\rm SH}=\theta_{\rm SH}\times \sigma \frac{\hbar}{e}, 
\end{equation}

where, $\sigma$ is the charge conductivity of the Ta thin films. We found the spin Hall conductivity, $\sigma_{\rm SH}$ to be around -2439~$(\hbar/e)~\Omega^{-1}$cm$^{-1}$ for low-resistivity Ta having mixed crystalline phase for the sample grown at a growth rate of 0.62 $\rm \AA$/s.
To our knowledge, no theoretical calculation exists for the polycrystalline mixed phase Ta films. 
However, first principle calculation show that the \textit{intrinsic} spin Hall conductivity of $\beta-$Ta is $-378~(\hbar/e)~\Omega^{-1}$cm$^{-1}$, while that of  $\alpha-$Ta is $-103~(\hbar/e)~\Omega^{-1}$cm$^{-1}$.~\cite{qu2018prb} Hence, the significantly higher value of $\sigma_{\rm SH}$ , that we obtain for the mixed phase Ta is likely caused by extrinsic mechanism.  The intrinsic and extrinsic contributions to the spin Hall
conductivity can be written in the following manner:\cite{tian2009prl,isasa2015prb,ramaswamypra2017,sagasta2018arXiv}

\begin{equation}\label{eq:scale}
\sigma_{\rm SH}=\sigma_{\rm SH}^{\rm int}+ \frac{\sigma_{\rm SH}^{\rm sj}\rho_{\rm Ta, 0}^{2}+\alpha_{ss}\rho_{\rm Ta, 0}}{\rho_{\rm Ta}^{2}},
\end{equation}
where, $\sigma_{\rm SH}^{\rm int}$ is intrinsic spin Hall conductivity, $\sigma_{\rm SH}^{\rm sj}$ is the spin Hall conductivity due to side jump mechanism, $\alpha_{ss}$ is the skew scattering angle and $\rho_{\rm Ta}$ is the longitudinal resistivity of Ta at room temperature, and $\rho_{\rm Ta, 0}$ is the residual resistivity of Ta. In Fig.~\ref{SHA}(b), we plot the measured $|\sigma_{\rm SH}|$ versus $\rho_{\rm Ta}$. The figure shows a strong dependence of $\sigma_{\rm SH}$  with $\rho_{\rm Ta}$ indicating that the $\sigma_{\rm SH}$ in these films is influenced by both the intrinsic and extrinsic contributions. In particular, for the mixed phase Ta films, $\sigma_{\rm SH}$ increases when $\rho_{\rm Ta}$ decreases as predicted by Eq.~(\ref{eq:scale}).  Assuming that $\rho_{\rm Ta, 0}$ is independent of crystalline phase, we expect $\sigma_{\rm SH} \propto 1/\rho_{\rm Ta}^{2}$, which is shown by the dashed line. The experimental behavior nearly follows this dependence except for the  $\alpha-$phase Ta. Thus, from Fig.~\ref{SHA} (b) one can conclude that the large spin Hall conductivity in the mixed phase Ta films is due to the extrinsic mechanism of spin Hall effect. Though a more detailed microscopic examination of samples is needed to find the exact origin of defects in the mixed phase Ta, the results do support the extrinsic mechanism of spin Hall effect in the Ta thin films. 



In summary, we have measured inverse spin Hall voltage (ISHE) in Py/Ta system by varying the crystalline phase of Ta using a co-planar wave-guide based broadband ferromagnetic resonance set-up. We demonstrate a strong correlation of measured spin mixing conductance and spin Hall conductivity with the crystalline phase of Ta thin films. We found a large spin Hall conductivity of $-2439~(\hbar/e)~\Omega^{-1}$cm$^{-1}$ for low-resistivity (68~$\mu\Omega$--cm) Ta having mixed crystalline phase, due to the extrinsic mechanism of spin Hall effect. The study is useful for the efficient manipulation of magnetization at the nanoscale as well as for explaining the spread in the values of spin Hall angle of Ta in literature.

The partial support from the Ministry of Human Resource Development under the IMPRINT program, the Department of Electronics and Information Technology (DeitY), and Department of Science and Technology under the Nanomission program are gratefully acknowledged. 
A.~K. acknowledges support from Council of Scientific and Industrial Research (CSIR), India.


%

\end{document}